# Coherent Spin Oscillations in a Disordered Magnet

S. Ghosh*, R. Parthasarathy*, T.F. Rosenbaum*, and G. Aeppli[†]

*The James Franck Institute and Department of Physics, The University of Chicago, Chicago, IL 60637

[†]NEC Research Institute, 4 Independence Way, Princeton, NJ 08540

Most materials freeze when cooled to sufficiently low temperature. We find that magnetic dipoles randomly distributed in a solid matrix condense into a spin liquid with spectral properties on cooling that are the diametric opposite of those for conventional glasses. Measurements of the non-linear magnetic dynamics in the low temperature liquid reveal the presence of coherent spin oscillations composed of hundreds of spins with lifetimes up to ten seconds. These excitations can be labeled by frequency, manipulated by the magnetic fields from a loop of wire, and permit the encoding of information at multiple frequencies simultaneously.

Magnetic solids offer arrays of quantum degrees of freedom, or spins, interacting with each other in a manner and strength ranging from the long-range ferromagnetism of iron and nickel to the nano-antiferromagnetism of vortices in high temperature superconductors. Unfortunately, there is a large barrier to exploiting quantum effects in magnetic solids: namely, the rarity of coherence effects that can be simply manipulated and observed *(1)*. In particular, it is difficult to create the magnetization oscillations corresponding to prepared superpositions of states, which are so straightforwardly created in liquid phase nuclear magnetic resonance (NMR) experiments. The "decoherence" for the solid magnets is generally attributed to disorder and to the coupling of the electronic spins to other degrees of freedom, such as nuclear spins, atomic motion, and conduction electrons. The present paper describes coherence effects in a magnet,

LiHo$_{0.045}$Y$_{0.955}$F$_4$, which is highly disordered, but does not suffer from coupling to either conduction electrons or to atomic motions because it is a strongly ionic insulator with the spins derived from small, non-overlapping electronic orbitals.

Pure LiHoF$_4$ is a ferromagnet because of its crystal structure and the dominant role of the magnetic dipole interaction *(2)* between the unpaired f electrons of the Ho$^{3+}$ ions. When non-magnetic Y$^{3+}$ ions are randomly substituted for the Ho$^{3+}$ ions, the fact that the anisotropic dipolar interaction can be antiferromagnetic as well as ferromagnetic begins to matter. The result is that with decreasing dipole concentration the ferromagnetism is first suppressed and then destroyed, to be replaced by conventional spin freezing *(3)*, analogous to the freezing of liquids into glasses. In a glass, barriers to relaxation proliferate with decreasing temperature and the system response slows and broadens. A surprising result of an early study *(4)* was that contrary to intuition as well as to theory *(5)*, further dilution of Ho by Y eliminated the glassy freezing in favor of a state that becomes progressively less glassy on cooling to temperatures as low as tens of mK. The "antiglass" state for LiHo$_{0.045}$Y$_{0.955}$F$_4$ is the host for the coherence effects that we have discovered.

We cooled a single crystal of LiHo$_{0.045}$Y$_{0.955}$F$_4$ with dimensions (1.0 x 0.5 x 0.5) cm$^3$ to mK temperatures using a helium dilution refrigerator. The energy scale is set by the nearest-neighbor dipole interaction strength, J$_{nn}$ = 1.2 K. Internal crystal fields force the randomly distributed Ho dipoles to point along the long crystalline c-axis, providing a one-dimensional Ising spin symmetry. We focus on how the sample magnetization responds to oscillating external magnetic fields along the c-axis of varying amplitudes h$_{ac}$ *(6)*. Ordinary disordered magnets respond to a changing external field via exponential relaxation *exp(-t/τ)*, where *τ* is a characteristic time. When transformed into the frequency domain, this corresponds to the Debye form *(7)*, $\chi(f) = \frac{\chi_o}{1 + 2\pi i f \tau}$. Glasses are typically described by a distribution of relaxation times, resulting in a response function that is the superposition of Debye forms for the times *τ* in the distribution. The outcome is a spectrum that becomes progressively wider on cooling *(3,8)*.

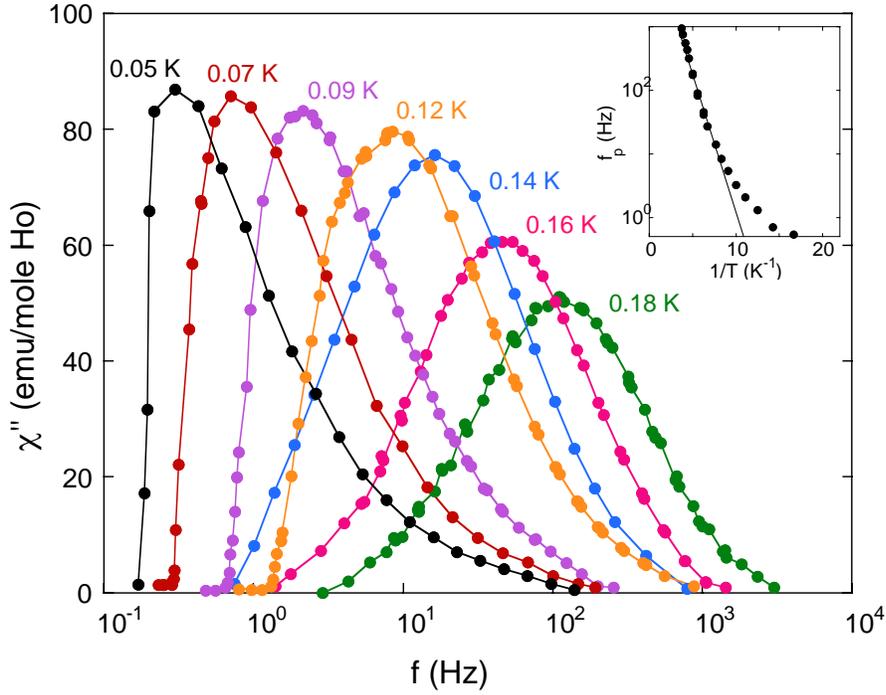

FIG. 1: Imaginary ($\chi''$) part of the magnetic susceptibility over five decades of frequency $f$ for the diluted, dipolar-coupled Ising magnet LiHo$_{0.045}$Y$_{0.955}$F$_4$. As expected, the dynamical response moves to lower frequency with decreasing temperature, but contrary to expectations for glassy systems, the spectral width narrows – and hence the barriers to spin relaxation diminish – as the disordered magnet is cooled. At the lowest temperatures the spectrum is narrower than Debye, indicating that classical relaxation of spins and spin clusters cannot describe the data. Lines are guides to the eye. Inset: Deviations from classical Arrhenius behavior for the peaks of the spectra (collected on cooling) emerge below T ~ 0.12 K.

We plot $\chi''(f)$ for LiHo$_{0.045}$Y$_{0.955}$F$_4$ over five decades in $f$ at seven different temperatures (Fig. 1). Spin relaxation becomes slower at low temperature, and the spectral response moves to lower $f$ with decreasing T. The typical frequency for spin reorientation, given by the peak in $\chi''(f)$, falls below 1 Hz for T < 0.090 K. Even while the overall dynamics slows, the spectrum continues to narrow down to the lowest T, opposite to the behavior observed in regular dielectric and magnetic glasses. The response function actually becomes narrower than Debye: 0.8 decades at T = 0.050 K as compared to the theoretical limit of 1.14 decades full width at half maximum for a single relaxation time. Moreover, the distribution of relaxation times becomes severely truncated at low frequencies (long times) as an apparent gap *(9,10)* opens in the spectrum below a cutoff *f* varying between 1 and 0.1 Hz on cooling from 0.120 to 0.050 K.

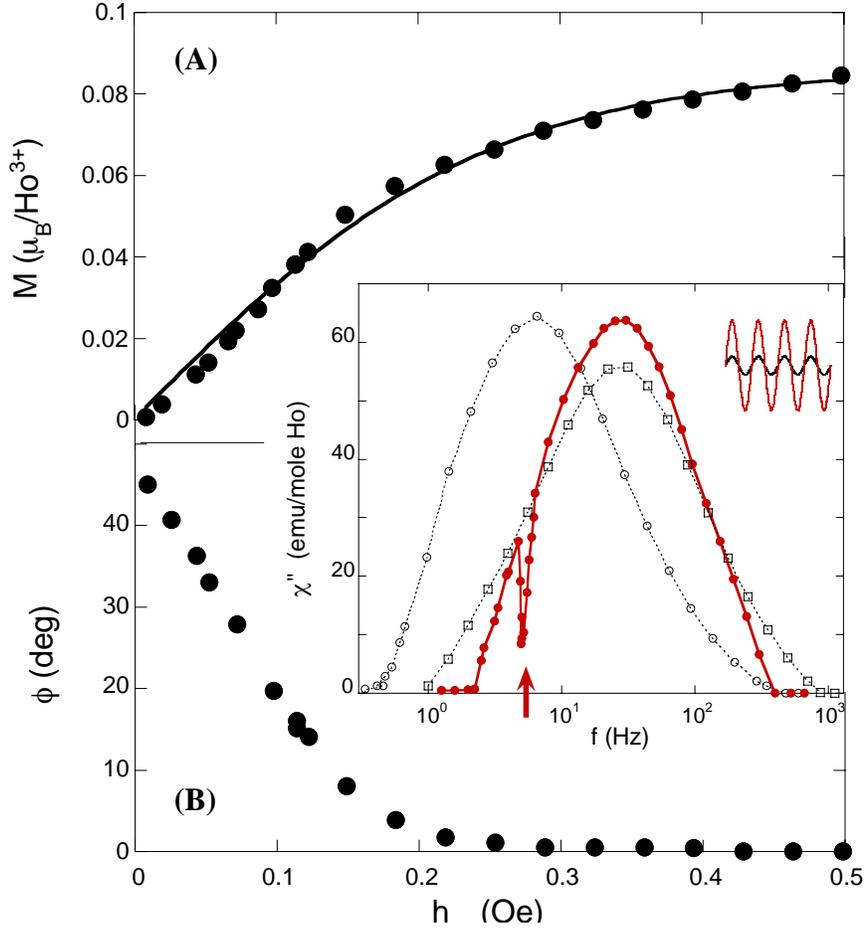

FIG. 2: (**A**) Amplitude and (**B**) phase lag of the magnetization oscillation appearing in response to a 5 Hz field oscillation at T = 0.110 K with excitation amplitude $h_{ac}$. The solid line is the Brillouin function, $M = Ngm_B s \cdot \tanh(ngm_B s h_{ac}/k_B T)$, for an s=1/2 Ising spin with a g-factor of 14. A least squares fit gives n = 260 for the number of Ho spins (each carrying $7m_B$) per cluster, with N = 7.0 x $10^{17}$ cm$^{-3}$ total spins responding. Inset: The imaginary part of the magnetic susceptibility at T = 0.110 K with (red) and without (black circles) a 0.2 Oe pump pulse at *f* = 5 Hz. The pump burns a hole in the spectrum at that frequency, indicating that the spectrum is composed of a discrete set of independent oscillators. Comparison to a spectrum at T = 0.150 K (black squares), with coincident $f_p$ but a larger intrinsic width, rules out simple heating due to the pump.

The sharpened spectra and the appearance of a gap imply that we are dealing with a set of oscillators rather than a distribution of relaxation times. An important consequence is that we should be able to address the oscillators individually via the technique of resonant hole burning, exactly as for optical systems. Hole burning consists of saturating an inhomogeneously broadened resonance line at its resonance frequency, bleaching the system response at that frequency *(11)*. The first signature of bleaching is saturation of the signal amplitude with excitation field. Fig. 2A demonstrates exactly such an effect for a 5 Hz excitation applied to our disordered magnet at T = 0.110 K. In addition, Fig. 2B shows that the phase lag between the signal and the excitation essentially vanishes on crossing between the linear, low-field and saturated, high-field regimes. Thus, dissipation decreases sharply on raising the excitation field.

The resonant character of the high-field, non-dissipative regime is established by simultaneously applying a high amplitude pump and a small amplitude probe along the Ising axis (Fig. 2, inset). The pump frequency is kept fixed while that of the probe is varied to obtain the spectrum. The pump has several significant effects. First, the spectrum's peak shifts from 6 Hz to 27 Hz. Second, while the spectral shape below the peak remains unchanged on the logarithmic-linear scale used to plot the data, it narrows above the peak. Furthermore, the spectrum is considerably narrower than that found when the temperature is raised to 0.150 K, to obtain (in the small probe amplitude limit) the same peak frequency, and we conclude that the pump is not simply heating the sample. Finally, and most dramatically, the pump amplitude of merely 0.2 Oe (five times the probe amplitude $h_{ac}$ = 0.04 Oe) at *f* = 5 Hz carves out a hole in $\chi''(f)$ that removes 75% of the original signal. No other portion of the spectrum is similarly affected. Varying the pump frequency burns holes of similar depth and width for any given *f* < $f_p$. A macroscopic number of spins defines any such independent state: integrating $\chi''(f)/f$ over *f* and normalizing by T yields ~ 7.5 x $10^{17}$ spins $cm^{-3}$ transparent to the probe, which is fully 1.5% of the total number accessible at T = 0.110 K.

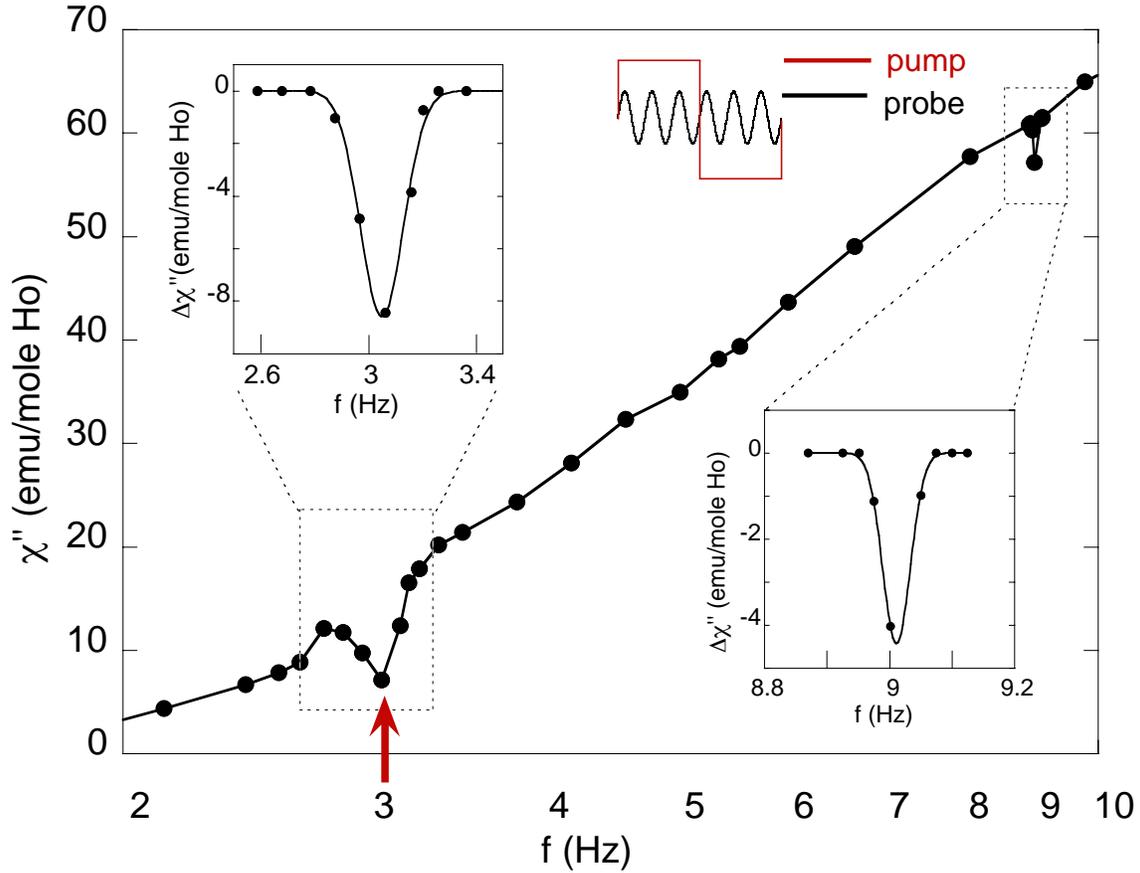

FIG. 3: Multiple holes burned into the spectrum using a square wave pump at $f = 3$ Hz (second Fourier component at $f = 9$ Hz), demonstrates the ability to encode simultaneously multiple bits of information. The hole at $f = 15$ Hz is not shown.

If the oscillators excited at one resonance frequency are independent from other oscillators in the antiglass, then it may be possible to burn holes in parallel at different frequencies. To this end, we substituted a square wave pump for the sinusoidal pump. By Fourier decomposition of the square wave, the substitution is equivalent to applying simultaneous sinusoidal pumps at $f_o$, $3f_o$, $5f_o$… with peak-to-peak amplitudes $a_o$, $a_o/3$, $a_o/5$ …, respectively. In the resultant spectrum (Fig. 3), not only is there a hole at 3 Hz, but there is a second one at 9 Hz, and the latter is about one-third as deep as the first, which is the ratio of excitation amplitudes in the Fourier expansion.

How long can the collective spin excitations maintain coherence against dissipation? We answer this question by looking for free induction decay after applying a pump signal and then shutting it off (Fig. 4A). The magnetization continues to oscillate at the pump frequency for times very much longer than the 0.2 s oscillation period of the pump. The extent of the free induction decay depends on temperature: the decay time rises from 4 s to 10 s on cooling from 0.125 to 0.070 K.

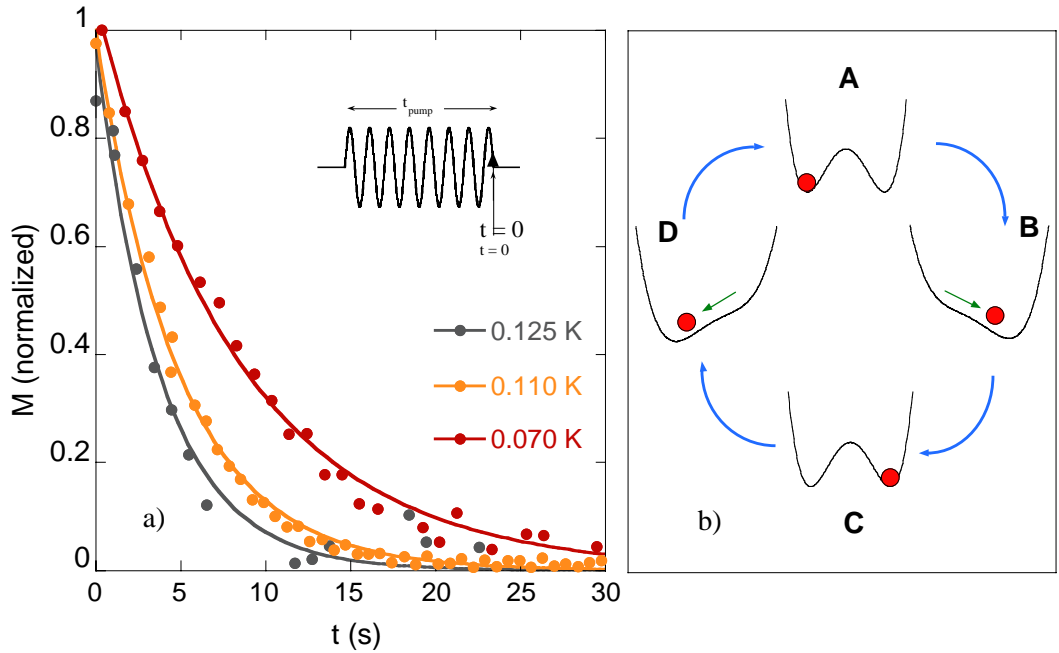

FIG. 4: (**A**) Persistent magnetization oscillations seen after running a 5 Hz pump with an amplitude of 0.2 Oe for 600 sec. The zero of time refers to when the pump is shut off, and the data are in the 5 Hz rotating frame in phase with the pump. As the sample is cooled, dissipation decreases and the oscillations persist for progressively longer times. (**B**) Schematic of spin clusters transiting between their up and down magnetic states at low temperatures during different parts of an ac drive cycle. The cluster magnetization is a degree of freedom subject to a potential which for zero applied field (A & C) contains two degenerate minima corresponding to up and down states. External fields tilt the potential, until at sufficiently high field only a single minimum survives, corresponding to magnetization along the field direction (B & D). An ac drive field produces an oscillating potential for the cluster magnetization, resulting in a variety of stochastic effects at low drive amplitudes *(19,20)* and saturation and phase locking effects at high amplitudes.

The coherent hole-burning and free induction decay of the magnetic excitations in LiHo$_{0.045}$Y$_{0.955}$F$_4$ appear to be derived from a collection of nearly independent oscillators embedded in a spin liquid rather than the relaxational modes responsible for the spectral depletion phenomena seen in conventional magnetic and dielectric solids *(12-14)*. What might these oscillators be? The most natural possibility is that they are magnetic clusters switching between states with up and down polarization. A cluster can be represented as a particle with a single coordinate – its magnetization – moving in a potential with two minima corresponding to the up and down polarized states (Fig 4B). For zero applied field, the minima will be separated by a barrier, and switching between the minima occurs at random intervals due to a combination of thermally activated hopping over and quantum tunneling through the barrier. In large fields the barrier disappears, and there is only one minimum corresponding to a fully polarized cluster. If the system is driven by a small ac magnetic field, the random barrier hopping and tunneling processes will result in dissipative dynamics, but for high drive amplitudes the motion will be deterministic over a large part of the oscillatory cycle. It is then describable in terms of Newtonian equations, and even can become dissipationless (Fig. 2B). The long-term ringing which we observe subsequently as free induction decay (Fig. 4A) is perhaps the most striking manifestation of the small dissipation for our spin clusters, and points to a large inertial term that is much larger than the dissipative term in the underlying equations of motion *(15)*.

The origin of nearly dissipationless, non-linear dynamics is no doubt linked to that of the antiglass phenomenon. Understanding LiHo$_{0.045}$Y$_{0.955}$F$_4$ must require an ingredient that is missing from the theory of randomly distributed classical dipoles, for which a glass transition is anticipated even in the limit of infinite dilution *(5)*. We suspect that the ingredient is quantum mechanics, which can play a role in the dilute system because the off-diagonal terms in the dipolar interaction can act as spin-flipping transverse fields. For disordered samples with sufficient Ho density to achieve ferromagnetism, these components of the dipolar interaction perpendicular to the Ising axis give rise to quantum mechanical ferromagnetic domain wall tunneling which survives even in the limit of zero external transverse field *(16)*. For our dilute crystals, the net outcome is neither a ferromagnetic nor a spin glass ground state *(17,18)*, but instead a subdivision of the system into clusters, whose conventional freezing is eventually limited by the quantum fluctuations brought about by the transverse fields from other clusters. The coherent oscillations that we observe could then be thought of as Rabi oscillations of the clusters in the weak transverse mixing fields.

We can use our data to see the crossover to quantum behavior and to measure the size of the clusters. The classical-quantum crossover is apparent from plotting (Fig. 1 inset) the frequency $f_p$ where the imaginary part of the magnetic response peaks against temperature. At high T a thermally-activated Arrhenius form, $f_p \sim exp(-\Delta/k_B T)$, fits the data with a barrier height $\Delta = 1.2$ K $\sim J_{nn}$. For T < 0.120 K, the sharp, gap-like cutoff appears in $\chi''(f)$ and there is a clear deviation from the Arrhenius law in the sense that the dynamics are too fast. We surmise that the dominant effect at low frequencies, drive amplitudes and temperatures becomes thermally-assisted quantum tunneling, with perhaps a role for stochastic synchronization effects *(19,20)* in defining the sharp, but nonetheless temperature-dependent, low frequency cutoff (Fig. 1).

The size of the clusters can be derived from the dependence of the magnetization on external magnetic field. The saturation of the magnetic response in Fig. 2A follows the familiar Brillouin form for Ising spins, M $\sim$ tanh(mh$_{ac}$/k$_B$T). Unlike a simple paramagnet, m here is not the magnetic moment of a single Ho ion *(21,22)*, but rather the total moment of the spins locked together in the cluster. Analysis of this non-linear response reveals that the clusters responsible for the saturation and hole burning effects at T = 0.110 K and $f$ = 5 Hz contain approximately 260 spins. This corresponds to cluster dimensions of approximately 6 Ho-Ho spacings on a side, and a probability of cluster membership of 1% for any given spin in the crystal, comparable to that deduced from the spectral weight carved out by the hole. A cluster of 260 spins each carrying 7$\mu_B$ and driven at h$_{ac}$ = 0.5 Oe sets an effective energy scale $\sim$ 0.13 K, comparable to the measuring temperature T, the onset temperature for deviations from Arrhenius behavior, and the temperature at which the spectral gap opens.

Beyond the implications for the problem of disordered magnets, our data demonstrate the ability to imprint phase coherent information in a chemically homogeneous bulk magnetic material using frequency as a label. This means that solid magnets may yet have a future in quantum information processing applications where coherent spin oscillations are actively manipulated to implement computations *(23)*. The necessary next step would involve entangling the states, a possibility that can be explored in Li(Ho,Y)F$_4$ because for this material an external transverse field readily produces quantum mixing *(16,24)*.

25. We have benefited greatly from discussions with J. Brooke, P. Chandra, S. Coppersmith, and S. Girvin. The work at the University of Chicago was supported primarily by the MRSEC Program of the National Science Foundation under Award No. DMR-9808595.